%Paper: hep-th/9407043
%From: RAGNISCO%40808@icil64.cilea.it
%Date: Thu, 7 Jul 1994 17:33:44 MET-DST

\magnification=\magstep1
\baselineskip = 18 true pt
\hsize =  17 true  cm
\vsize =  23 true cm
\centerline {{\bf DYNAMICAL R-MATRICES FOR INTEGRABLE MAPS}}
\vskip 1 true cm
\centerline { O.Ragnisco}
\centerline {{\it Dipartimento di Fisica E.Amaldi, III Universit\`a di Roma }}
\centerline {{\it and}}
\centerline {{\it Istituto Nazionale di Fisica Nucleare, Sezione di Roma,
P.le A.Moro 2, 00185 Roma,Italia} }
                        \vskip 1 true cm

\vskip 2 true cm
\vskip 1 true cm
\noindent {\bf Abstract}

\noindent
The integrability of two symplectic maps, that can be considered as
discrete-time analogs of the Garnier and Neumann systems is established
in the framework of the $r$-matrix approach, starting from their
Lax representation. In contrast with the
continuous case, the $r$-matrix for such discrete systems
turns out to be of dynamical type; remarkably, the induced Poisson
structure  appears as a linear combination of compatible ``more
elementary" Poisson structures. It is also shown that
the Lax matrix naturally leads
to define separation variables, whose discrete and continuous
dynamics is investigated.
\vfil\eject

\noindent {\bf 1. Introduction}

In a number of recent papers, it has  been shown that the well-established
techniques devised to construct integrable finite-dimensional
hamiltonian systems out of  integrable
hierarchies of nonlinear evolution equations (often denoted
``soliton hierarchies") [1] can also be applied in a discrete context,
naturally
leading to integrable maps [2-5].
 As remarkable results, I would like to quote the construction of discrete-time
versions of the Toda-lattice [6] and of the Calogero-Moser system [7].

In this context, the author has shown [5] that, starting from the so-called
``Toda Hierarchy with Sources", one readily obtains integrable maps equipped
with a Lax pair. In a further paper [8], the connection of these maps with
the stationary Toda flows and with the finite gap sector of the solution
manifold
of the Toda hierarchy has been rigorously established.

%I have to mention that a different  route to achieve the previous goal
%has been followed by Veselov [5] and by Veselov and Moser [6], who
%constructed discrete-time  integrable versions of spin chains, Euler tops
%and Neumann systems.

In the  present paper, we investigate the algebraic structure underlying
two of  such integrable maps, namely the Discrete Garnier (DG) [3] and
the Discrete Neumann (DN) [4]  systems, in terms of the ``classical
$r$-matrix" formalism as generalized in a fundamental
paper by Babelon and Viallet [9].
Surprisingly, we find that the associated $r$-matrices are of dynamical type,
in contrast with those pertaining to the corresponding continuous systems
[10,11],
and moreover appear as a linear combination of ``more elementary" $r$-matrices,
so that the associated Poisson structure is itself a linear combination
of compatible Poisson structures.

We recall that $r$-matrices of dynamical type have recently been discovered
for an extremely important class of integrable finite-dimensional
continuous time systems, namely the Calogero-Moser class [12].

A remarkable property of the maps under scrutiny is their separability
in terms of ``roots variables" naturally provided by the Lax representation,
in full analogy with the continuous systems [11].

The DG and DN systems and their Lax representation are tersely
 reviewed in Section 2.

In Section 3 we derive the associated $r$-matrices, while in Section 4 we
define separation variables and investigate the corresponding dynamics.

A few (possibly) interesting open problems are mentioned in Section 5.
\vskip 1 true cm
\noindent {\bf{2.~Two Integrable Maps and their Lax Representation}}

The  DG system is given by the following Lagrangean map:

$$<\Psi_n,\Psi_{n-1}>\Psi_{n-1}+<\Psi_n,\Psi_n>\Psi_n+<\Psi_n,\Psi_{n+1}>
\Psi_{n+1}=\Lambda\Psi_n\eqno(2.1)$$

\noindent
In (2.1), $n\in {\bf{Z}}$, $\Psi_n$ is an ${\bf{R}}^N$-vector of components
$\Psi_n^{(j)}$, $\Lambda$ is a diagonal matrix with distinct entries
$(\lambda_1,\cdots,\lambda_N)$ and the symbol $<\cdot,\cdot>$ denotes the
usual euclidean inner product in ${\bf{R}}^N$.

As explained in [5], the system (2.1) arises by taking $N$ replicas of
the Toda-lattice spectral problem [13]:

$$a_{n-1}\Psi_{n-1}^{(j)} + b_n\Psi_n^{(j)} +a_n\Psi_{n+1}^{(j)} =
\lambda_j\Psi_n^{(j)}. \eqno(2.2)$$

\noindent
corresponding to $N$ different values of the spectral parameter $\lambda$, and
by restricting the dynamical variables $(a_n, ~b_n)$ to the invariant manifold:

$$a_n=2<\Psi_n,\Psi_{n+1}>;~~~~b_n=<\Psi_n,\Psi_n>\eqno(2.3)$$

\noindent
The corresponding Lagrange function reads:

$${\cal{L}} (x,y)~=~<x,y>^2+{1\over 2}<x,x>-<x,\Lambda x>$$

\noindent
where $x=\Psi_n,~y=\Psi_{n+1}$.

\noindent
Canonical variables $q,p$ can be introduced in the standard way [2]:

$$q=\Psi_n;~~~p={\partial {\cal{L}}\over \partial
y}|_{(x=\Psi_{n-1},y=\Psi_n)}=
{\Psi_{n-1}\over <\Psi_n,\Psi_{n-1}>}\eqno(2.4)
$$

\noindent
thus allowing to recast (2.1) in the form of the following   map:

$$q^\prime~=~(\Lambda - <q,q>q - p)/g\eqno(2.5a)$$

$$p^\prime~=~gq\eqno(2.5b)$$

\noindent
with $g^2=<q,\Lambda q>-<q,q>^2-<p,q>$. The map (2.5) is symplectic for the
standard symplectic
form $dq\wedge dp$.
Exploiting the procedure outlined in [5], the map (2.5) can be represented
in the following {\it{discrete}} Lax form:

$$L=AL^\prime A^{-1}\eqno(2.6a)$$

\noindent
where $L,A$ are $2\times 2$ matrices, given by:

$$L=\pmatrix{-(\lambda/2 + <p,R_\lambda q>)&{\sqrt {<p,q>}}
(1+<p,R_\lambda p>/<p,q>\cr -{\sqrt {<p,q>}}(1+<q,R_\lambda q>)&
\lambda/2 + <p,R_\lambda q>\cr}\eqno(2.6b)$$

$$A=\pmatrix{(\lambda-<q,q>)/{\sqrt {<p,q>}}&-
{\sqrt {<p^\prime,q^\prime>}}/{\sqrt {<p,q>}}\cr 1 & 0\cr}\eqno(2.6c)$$

\noindent
where
$$R_\lambda=(\lambda I -\Lambda)^{-1}$$

\noindent
The meromorphic invariant function $\Delta(\lambda)\equiv det[L(\lambda)]$
is  the generating function of the conserved quantities:
$$\Delta(\lambda)=
-\lambda^2/4+\sum_{j=1}^N{I_j\over \lambda-\lambda_j}\eqno(2.7a)$$

$$I_j=\sum_{k\ne j}{(p_jq_k-p_kq_j)^2\over\lambda_k-\lambda_j}+p_j^2+
<p,q>q_j^2-\lambda_jp_jq_j\eqno(2.7b)$$
\noindent
and an explicit computation shows that they are mutually in involution

$$\{I_j,I_k\}=0,$$

\noindent
thus entailing the complete integrability of the map (2.5).

Similar considerations hold for the DN system [4], described by the map:

$${\Psi_{n-1} \over 2<\Psi_{n-1},\Psi_n> } +b_n \Psi_n + {\Psi_{n+1}
 \over 2<\Psi_{n+1},\Psi_n> }=\Lambda \Psi_n
\eqno(2.8)$$

\noindent
where the discrete motion is constrained on the unit sphere $S^N$:
$<\Psi_n,\Psi_n>=1$, and accordingly the Lagrange multiplier $b_n$
is determined to be: $b_n=<\Psi_n,\Lambda \Psi_n>-1$.

\noindent
Eq.(2.8) is a Lagrangean map for the Lagrange function:

$${\cal{L}} (x,y)=log<x,y>~-~<x,\Lambda x>$$
\noindent
on $S^N$.

\noindent
In terms of the hamiltonian variables $q,p$, defined as in (2.4)
by :
$$q=\Psi_n;~~~p={\partial {\cal{ L}} \over \partial
y}|_{x=\Psi_{n-1},y=\Psi_n}=
{\Psi_{n-1}\over 2<\Psi_n,\Psi_{n-1}>}$$
eq.(2.8) becomes the map:
$$ p^\prime=q||(\Lambda-b)q-p||~~;~~
 q^\prime={(\Lambda-b)q-p\over ||(\Lambda-b)q-p||}\eqno(2.9a)$$
$$||q||=1,~~~<q,p>={1\over  2}\eqno(2.9b)$$

\noindent
which is symplectic for the Poisson brackets:

$$\{q,q\}=0;~~\{q,p\}=I - q\otimes q;~~\{p,p\}=p\wedge q
\eqno(2.9c)$$

\noindent
It enjoys the Lax representation (2.6a), with:

$$L=\pmatrix{1/2~+~<p,R_\lambda q>  &  -<p,R_\Lambda p>/\Vert p \Vert\cr
             <q,R_\Lambda q>/\Vert p \Vert & -1/2~-~<p,R_\lambda q>\cr}
\eqno(2.10a)$$

$$A=\pmatrix{(\lambda-p)/\Vert p \Vert & -\Vert p^\prime \Vert/\Vert p \Vert\cr
                         1             &               0
\cr}
\eqno(2.10b)$$

\noindent
Again, the invariant function $\Delta(\lambda)\equiv det[L(\lambda)]$ is the
generating function of the integrals of motion:

$$\Delta(\lambda)=\sum_j {I_j\over \lambda-\lambda_j}\eqno(2.11a)$$

\noindent
with

$$I_j=\sum_{k\ne j}{(p_jq_k-p_kq_j)^2\over\lambda_k-\lambda_j}-q_jp_j
\eqno(2.11b)$$

\noindent
The conserved quantities $I_j$ have been shown in [4] to be in involution
for the Poisson brackets (2.9c).
\vskip 1 true cm

\noindent
{\bf{3. r-Matrix Formulation}}
\medskip

As it has been shown for the first time in [9], whenever a hamiltonian system
is associated with a Lax matrix whose eigenvalues are in involution for a
given Poisson bracket, it enjoys an $r$-matrix representation of the form:

$$\{L_1,L_2\}~=~[r_{12},L_1]-[r_{21},L_2]\eqno(3.1)$$

\noindent
We have used the standard notations:

$$L_1\equiv L\otimes {\bf{1}}~=~\sum_i L_i e^i\otimes {\bf{1}}$$

$$L_2\equiv{\bf{1}}\otimes L~=~\sum_iL_i {\bf{1}}\otimes e^i$$

$$r_{12}=\sum_{i,k}r_{ik} e^i\otimes e^k~~~~~r_{21}=
\sum_{i,k}r_{ki} e^i\otimes e^k$$

\noindent
where $\{e^i\}_{i=1}^M$ is a basis for the matrix Lie-algebra which $L$ belongs
to, and, in general, the coefficients $r_{ik}$ will be functions on the phase
space ({\it{dynamical r-matrix}}).

For both the systems introduced in Sec.2, the Lie algebra is obviously
$sl(2)$, and we will use the Cartan-Weil basis:

$$\sigma^3=\pmatrix{1 & 0\cr 0 & -1\cr};~~\sigma^+=\pmatrix{0 & 1\cr 0 & 0\cr};
{}~~\sigma^-=\pmatrix{0 & 0\cr 1 & 0\cr}\eqno(3.2a)$$

\noindent
rather than the Pauli basis:

$$\sigma^1=\sigma^+ + \sigma^-;~~\sigma^2= i(\sigma^- - \sigma^+);~~ \sigma^3
\eqno(3.2b)$$

As a matter of fact, since $L$ has an additional (rational) dependence upon
the spectral parameter $\lambda$, it has to be regarded as an element of the
loop algebra ${\cal{G}} =sl(2)\otimes C(\lambda, \lambda^{-1})$, namely as a
formal Laurent series in $\lambda$ with coefficients in $sl(2)$:

$$L(\lambda)=\sum_{k=-\infty}^p L_k\lambda^k\eqno(3.3)$$

\noindent
Incidentally, we notice that for matrices (2.6b), (2.10a), the series (3.3)
is actually uniformly convergent in any compact subset of the annulus
$max_j \vert\lambda_j \vert < \vert \lambda \vert <\infty$.

\noindent
The trace form on $\cal G$, given by:

$$(L)~\equiv~res~ tr~L(\lambda)~~=tr~L_{-1}\eqno(3.4)$$

\noindent
allows to identify $\cal G$ with its dual (the space of linear functions
on $\cal G$), and to consider the $r-$matrix as an endomorphism $R$ on $\cal
G$,
rather than as an element of $\cal G \otimes \cal G$. In such a ``dual"
picture,
eq.(3.1) induces the following Poisson bracket between two functions on
$\cal G$:

$$\{f,g\}_L~=~(L,[df,dg]_R)\eqno(3.5a)$$

\noindent
with

$$[X,Y]_R=[X,R(Y)]+[R(Y),X]~~~(X,Y\in \cal G)\eqno(3.5b)$$

\noindent
and:

$$R(X)~=~\sum_{j,k}r_{jk}e^{(i)}(e^{(k)},X)\eqno(3.5c)$$

\noindent
As is well known [9], eq.(3.1) implies involutivity of the invariants of the
matrix $L$.

\noindent
In the dual picture, it is convenient to look at the functions:

$$f_k~=~res~{1\over 2}tr~\lambda^kL^2~=~{1\over 2}(L,\lambda^kL)\eqno(3.6a)$$

\noindent
{}From (3.5):

$$\{f_k,f_j\}~=~(L,[df_k,df_j]_R)~=~(L,[R(\lambda^kL),\lambda^jL])+
{}~(k\leftrightarrow j)=0\eqno(3.6b)$$

\noindent
In our concrete cases, where $L$-matrices are given by (2.6b) and (2.10a),
the functions $f_k$ are related to the invariants $I_j$ (2.7b),(2.11b)
by the  formula:

$$f_k=\sum_j \lambda_j^k I_j\eqno(3.7a)$$

We can summarize the previous result by the following Theorem:
\medskip
\noindent
{\bf{ Theorem}} 1.

\noindent
{\it{The  hamiltonian flows of the functions $f_k$ correspond to completely
integrable continuous-time hamiltonian systems, endowed with the Lax
representation}}:

$${\partial L\over \partial t_k}~=~[L,R(\lambda^kL)]$$
\medskip
\noindent
The proof of the above Theorem is straightforward. The complete integrability
follows from (3.6b). As  for the Lax representations, we have, by definition:

$${\partial L\over \partial t_k} := \{L, f_k\}\eqno(3.7b)$$

\noindent
On the other hand, for any basis element $\sigma^l$, and for any integer
$j$, it holds:

$$\{(\sigma^l,\lambda^jL),f_k\}=(\lambda^j\sigma^j,[L,R(\lambda^kL)])$$

\noindent
namely:
$$tr~ \sigma^l{\partial (res~\lambda^jL)\over \partial t_k}~=~
tr~ \sigma^l~(res~\lambda^j[L,R(\lambda^kL)])$$
\noindent
and thus

$$res~\lambda^j({\partial L\over \partial t_k}-[L,R(\lambda^kL)])=0$$
\noindent
whence  the result. $\diamond$
\medskip
It is worth to notice that there is an  intimate relation between an
integrable map and the hamiltonian flows of its invariant functions;
namely it holds the following Proposition:
\medskip
{\bf{Proposition I}}

\noindent
 {\it{An integrable map is a Backlund transformations for the hamiltonian
flows of its invariants.}}

\noindent
The proof of the above assertion is trivial: let us denote by $x$ a point
in the phase space $M=({\bf{R}}^{2N},\omega)$, and by $x^\prime=\Phi (x)$
a symplectic map possessing the $N$ invariant functions ${\cal {F}}_j$, in
involution with respect to the Poisson bracket induced by  $\omega$.

\noindent
Let:

$$K^{(j)}\equiv {\partial x\over \partial t_j}=\{x,{\cal {F}}_j\}\vert_x$$

\noindent
then:

$${\partial x^\prime\over \partial t_j}=\dot\Phi (x)\cdot
{\partial x\over \partial t_j}~=~\dot\Phi (x)\cdot \{x,{\cal {F}}_j(x)\}~=
{}~\{\Phi(x),{\cal {F}}_j(x)\};$$

\noindent
but ${\cal {F}}_j$ is  an invariant function for $\Phi$, and thus:

$${\partial x^\prime\over \partial t_j}=\{\Phi(x),{\cal {F}}_j(\Phi(x))\}$$

\noindent
and finally, since $\Phi$ is symplectic:

$${\partial x^\prime\over \partial
t_j}=\{x,{\cal{F}}_j\}\vert_{x^\prime=\Phi(x)}$$

\noindent
i.e. $\Phi$ maps solutions into solutions. $\diamond$
\medskip
We now proceed to the explicit evaluation of the $ r$ matrices.
\medskip
\noindent
1) DG {\it{system}}.

By direct calculation, we have:

$$
\{L_3(\lambda),L_3(\mu)\}=0$$

$$\{L_3(\lambda),L_\pm(\mu)\}=\pm{2\over
\lambda-\mu}[L_\pm(\lambda)-L_\pm(\mu)]
$$
$$\{L_\pm(\lambda),L_\pm(\mu)\}=
-{1\over{\sqrt{ <p,q>}}}[L_\pm(\lambda)-L_\pm(\mu)]$$
$$\{L_\pm(\lambda),L_\mp(\mu)\}=\pm{4\over
\lambda-\mu}[L_3(\lambda)-L_\pm(\mu)]
+{1\over{\sqrt {<p,q>}}}[L_\pm(\lambda)-L_\mp(\mu)]\eqno(3.8)$$

\noindent
The above formulas already  show the appearance of the ``dynamical term"
${1\over{\sqrt {<p,q>}}}$. They entail the following expression for the Poisson
bracket $\{L_1,L_2\}$:

$$\{L_1(\lambda),L_2(\mu)\}=$$
$$=-{1\over \lambda-\mu}[\Pi,L_1(\lambda)+L_2(\mu)]
+{1\over 2{\sqrt {<p,q>}}}([\sigma^3\otimes(\sigma^--\sigma^+),L_1(\lambda)]-
[(\sigma^--\sigma^+)\otimes\sigma^3,L_2(\mu)])\eqno(3.9)$$

\noindent
where $\Pi$ is the usual permutation operator:

$$\Pi=\sum_i \sigma^i\otimes \sigma^i$$

\noindent
Comparing with the general formula (3.1), we have the final  expression for
the $r-$matrix:

$$r_{12}(\lambda,\mu)=-{1\over \lambda-\mu}\Pi+
{1\over 2{\sqrt {<p,q>}}}\sigma^3\otimes(\sigma^--\sigma^+)
\eqno(3.10)$$

\noindent
To derive the endomorphism $R(X)$ (3.5c) that takes part in the Poisson bracket
(3.5a), first of all we recall [10] that the term ${1\over \lambda-\mu}\Pi$
dualizes into the difference of the projectors $\cal P_+$ and $\cal P_-$
on positive and (strictly) negative $\lambda$-powers:

$${\cal {P}}_+(X)~=~\sum_{k\ge 0}X_k\lambda^k$$

$$ {\cal{P}}_-(X)~=~\sum_{k<0}X_k\lambda^k$$

\noindent
As  for the dynamical term, we note that:

$$res\lambda^{-1} L~=~\pmatrix{0    &{ \sqrt {<p,q>}}\cr
                   -{\sqrt {<p,q>}} &  0 \cr}$$

\noindent
whence it  follows:

$${\sqrt{ <p,q>}}=res {1\over 2}\lambda^{-1}(L_+-L_-)~=+{1\over
2}(L,\lambda^{-1}
(\sigma^--\sigma^+))$$

\noindent
so that:

$$R(X)=X_--X_+ +{1\over (L,\lambda^{-1}
(\sigma^--\sigma^+))}\sigma_3(\sigma^--\sigma^+,X)\eqno(3.11)$$

\noindent
It might be of some interest looking at the role of the dynamical part
of the $r-$matrix in the Jacobi identity, that we write down both
in the tensor picture and in the dual picture:

$$[L_1,[r_{12},r_{13}]+[r_{12},r_{23}]+[r_{32},r_{13}]]+
[L_1,\{L_2,r_{13}\}-\{L_3,r_{12}\}]~+~cyclic ~perm.=0
\eqno(3.12a)$$

$$(L,[X,{\cal{B}} (Y,Z) + \{(L,Y),R(Z)\}- \{(L,Z),R(Y)\}])~+~cyclic~ perm.~=~0
\eqno(3.12b)$$

\noindent
In (3.12a) all quantities are understood to belong to $\cal G\otimes
\cal G\otimes \cal G$ and, as usual, the subscript denotes the space
which the corresponding tensor acts on nontrivially:

$$L_1~=L\otimes{\bf{1}}\otimes{\bf{1}},~~r_{12}=
\sum_{jk}r_{jk}e^i\otimes e^k \otimes {\bf{1}}, etc.$$

\noindent
In (3.12b), we have  shortly denoted by $\cal B(\cdot,\cdot)$ the
usual Yang-Baxter term, namely:

$${\cal{B}}(X,Y)=[R(X),R(Y)]-R([X,Y]_R)$$

For  the  sake of  simplicity, we  focus our attention on eq.(3.12a),
and write for a  moment:

$$r_{12}=r_{12}^{(c)}+r_{12}^{(d)}$$

\noindent
where $r_{12}^{(c)}$ stands for the constant part $-{1\over \lambda-\mu}\Pi$
and $r_{12}^{(d)}$ stands for the dynamical part
${1\over 2{\sqrt {<p,q>}}}\sigma^3\otimes(\sigma^--\sigma^+)$.

\noindent
We see the following:

\noindent
(i) As is well known $r^{(c)}$ satisfies the Yang-Baxter equation:

$$[r_{12},r_{13}]+[r_{12},r_{23}]+[r_{32},r_{13}]=0$$

\noindent
(ii) The mixed terms, containing both $r^{(c)}$ and $r^{(d)}$ vanish.

\noindent
(iii) The  dynamical part $r^{(d)}$ yield both quadratic and cubic terms
in $1\over {\sqrt {<p,q>}}$. The quadratic terms in the first commutator of
(3.12a) cancel with the quadratic  terms in the second commutator, while the
cubic
terms, appearing just in the second commutator, cancel  among themselves
due to  cyclic  permutations.

\noindent
We  point out that property (ii) entails that the Jacobi identity
 equation (3.12a) splits into two equations, involving separately
$r^{(c)}$ and $r^{(d)}$, which are both satisfied. In terms of
Poisson brackets, this means that for the DG system the Poisson bracket
(3.1) (or (3.5a)) is actually the sum of two {\it{compatible}}
Poisson brackets, generated by $r^{(c)}$ and $r^{(d)}$ respectively.

\medskip
\noindent
(2) DN {\it{system}}

The Poisson brackets between the elements of the Lax
matrix now read:

$$\{L_3(\lambda),L_3(\mu)\}=0$$

$$\{L_3(\lambda),L_\pm(\mu)\}=\pm{2\over
\lambda-\mu}[L_\pm(\lambda)-L_\pm(\mu)]
+{1\over \Vert p \Vert}L_\pm(\mu)(L_\pm(\lambda)-L_\mp(\lambda));$$
$$\{L_\pm(\lambda),L_\pm(\mu)\}={1\over \Vert p
\Vert}[L_\pm(\lambda)-L_\pm(\mu)]
+{2\over \Vert p \Vert}(L_3(\lambda)L_\pm(\mu)-L_3(\mu)L_\pm(\lambda));$$
$$\{L_\pm(\lambda),L_\mp(\mu)\}=$$
$$=\pm{4\over \lambda - \mu}[L_3(\lambda)-L_3(\mu)]
+{1\over \Vert p \Vert}[L_\pm(\mu)-L_\mp(\lambda)]+{2\over \Vert p \Vert}
[L_\mp(\lambda)L_3(\mu)-L\pm(\mu)L_3(\lambda)]\eqno(3.13)$$
Comparing with (3.7a), we notice that linear terms in $L$ are the same, up
to the substitution ${\sqrt {<p,q>}} \leftrightarrow \Vert p \Vert$;
consequently,
the $r-$matrix will be of the form:

$$r_{12}=-{\Pi\over \lambda -\mu}+{1\over 2\Vert p \Vert}\sigma^3\otimes
(\sigma^- - \sigma^+)+ r_{12}^{(d_2)}\eqno(3.14a)$$

\noindent
where $r_{12}^{(d_2)}$ takes care of quadratic terms in $L$ in (3.13). Skipping
out
the computational details, we report the result:

$$r_{12}^{(d_2)}(\lambda ,\mu)={1\over 2\Vert p
\Vert}(\sigma^++\sigma^-)\otimes
[\sigma^3,L(\mu)]\eqno(3.14b)$$

\noindent
and correspondingly:
$$r_{21}^{(d_2)}(\lambda ,\mu)={1\over 2\Vert p \Vert}
[\sigma^3,L(\lambda)] \otimes(\sigma^++\sigma^-)\eqno(3.14b)$$

\noindent
It is also possible to write the dynamical term in an invariant form.
In fact we have

$$\Vert p \Vert=-res~tr~ L\sigma^-~=-(L,\sigma^-)$$

\noindent
Consequently, the $r-$matrix (3.14) dualizes to:

$$R(X)=X_--X_+-{1\over 2(L,\sigma^-)}\sigma^3(\sigma^--\sigma^+,X)
-{1\over 2(L,\sigma^-)}(\sigma^++\sigma^-)([\sigma^3,L],X)\eqno(3.15)$$
\noindent
The check of the Jacobi identity is now considerably more involved than in the
DG case. However, the same remarkable phenomenon occurs :
namely, the Jacobi identity equations decouples in three equations individually
satisfied by $r^{(c)},~r^{(d_1)},~r^{(d_2)}$, thus entailing that the
Poisson structure characterizing DN is actually the sum of {\it{three
compatible
 Poisson structures}} engendered by $r^{c)},~r^{(d_1)},~r^{(d_2)}$. It is
perhaps
worthwwhile to point out that the dynamical nature of $r_{12}^{(d_2)}$, namely
the
presence of the $\Vert p \Vert ^{-1}$ factor is indeed essential: the tensor
$[\sigma^3,L(\lambda)] \otimes(\sigma^++\sigma^-)$ alone is {\it{not}} an
$r-$matrix.
\vskip 1 true cm

\noindent
{\bf{4.Separability}}
\medskip

In this Section we will show that the integrable maps denoted as DG and DN,
as well as the integrable continuous flows (3.7b) associated with them
enjoy the classical property of separability. Namely, we will show that there
exists a set of canonical coordinates $(\mu_j,\pi_j)_{j=1}^N$ such that:

$$\pi_j~=~{\partial W\over  \partial \mu_j}\eqno(4.1)$$

\noindent
where the function $W_j$ depend just upon the variable $\mu_j$ and the
integrals
of motion:

$$W_j=W_j(\mu_j,\{I_k\}_{k=1}^N)$$

\noindent
Paraphrasing, with slight modifications, the derivation presented in
[11] we will introduce the variables $\mu_j$ as the zeroes of suitable
polynomials taking part in the Lax matrix.

To this aim we notice that the $L_-$ element of the Lax matrices (2.6b),
(2.10a) can be written as:

$$L_-=\alpha {P(\lambda)\over Q(\lambda)}\eqno(4.2a)$$

\noindent
where
$$\alpha = -{\sqrt {<p,q>}}~~ (DG);~~~~\alpha=\Vert p \Vert~~(DN )
\eqno(4.2b)$$

\noindent
and:
$$
Q(\lambda)=\prod_{j=1}^N(\lambda-\lambda_j)\eqno(4.3a)$$
\noindent
 $P(\lambda)$  being  a monic polynomial of degree $N$ for the DG system,
of degree $N-1$ for the DN system, whose {\it{real}} zeroes, which are
functions
of the canonical variables $q,p$, we denote by $\mu_j$:
$$P(\lambda)=\prod(\lambda-\mu_j)\eqno(4.3b)$$
\noindent
Following [11] we set:

$$\pi_j=L_3(\mu_j)\eqno(4.4)$$
\noindent
i.e.:
$$\pi_j=-({\mu_j\over
2}+\sum_{k=1}^N{p_kq_k\over\mu_j-\lambda_k})~~~j=1,\cdots,N
{}~~(DG)$$
$$\pi_j={1\over 2}+\sum_{k=1}^{N}{p_kq_k\over\mu_j-\lambda_k}~~~j=1,\cdots,N-1
{}~~(DN).$$
\noindent
Then, by direct calculation, starting from formulas
(3.8), (3.9), and paraphrasing again [11]  we can establish the following
theorem:

{\bf{Theorem 2}}:
\noindent
{\it{The variables $\pi_j,\mu_j$ are canonically conjugated. Moreover,
they are separation variables.}}

\noindent
The separation equations are obviously given by:
$$\pi_j^2=\Delta(\mu_j)\eqno(4.5)$$
\noindent
and consequently  the  functions $W_j$ of formula (4.1) read:

$$W_j=\int^{\mu_j}d\lambda \sqrt\Delta(\lambda)\eqno(4.6)$$

\noindent
We shall now look at the continuous flows of the invariants $f_k$
taking part in Theorem 1. From the Lax equation:

$${\partial L\over \partial t_k}~=~[L,R(\lambda^kL)]\eqno(4.7)$$

\noindent
we deduce:
$${\partial L_-\over \partial t_k}=2L_-M_3^{(k)}-2M_-^{(k)}L_3$$

\noindent
whence:
$${\partial L_-\over \partial t_k}\vert_{\lambda=\mu_r}=
-2M_-^{(k)}(\mu_r)\pi_r\eqno(4.8)$$

\noindent
It is easily  seen that, both in the DG and in the DN case, only the constant
part of the r-matrix contributes to $M_-^{(k)}(\mu_r)$. Indeed, a direct
and simple calculation leads to the formula (here and in the following a
superscript dot denotes differentiation: e.g. $\dot Q (\lambda_k) \equiv
{\partial Q\over \partial \lambda}\vert_{\lambda=\lambda_s}$):

$$M_-^{(k)}(\mu_r)=2\alpha\sum_s {\lambda_s^k\over
\mu_r-\lambda_s}{P(\lambda_s)
\over \dot Q (\lambda_s)}\eqno(4.9)$$

\noindent
taking into account that:
$${\partial L_-\over \partial t_k}\vert_{\lambda=\mu_r}=-\alpha
{\dot P (\mu_r)\over Q(\mu_r)}{\partial \mu_r\over \partial t_k}$$

\noindent
we get from (4.8):

$${\partial \mu_r\over \partial t_k}=
{4Q(\mu_r)\over \dot P (\mu_r)}\sum_s {\lambda_s^k\over
\mu_r-\lambda_s}{P(\lambda_s)
\over \dot Q (\lambda_s)}\eqno(4.10)$$

\noindent
Recalling the so-called Lagrange interpolation formula, which is a plane
consequence of the residues theorem:

$$\sum_r{\mu_r^l\over \dot P (\mu_r)(\mu_r-\lambda_s)}=
-{\lambda^l\over P(\lambda_s)}~~~~~~~(l\le deg(P))\eqno(4.11)$$

\noindent
we can write:

$$\sum_r{\mu_r\over Q(\mu_r)\sqrt \Delta(\mu_r)}{\partial \mu_r^l\over \partial
 t_k}=
-\sum_s{\lambda_s^{k+l}\over \dot Q (\lambda_s)}\eqno(4.12)$$

\noindent
Eq.(4.12) can be immediately integrated, yielding  to the Jacobi
inversion problem (solvable in terms of Riemann $\Theta$ functions):

$$\sum_r \int_{\mu_r(t_k^{(0)})}^{\mu_r(t_k)}d\lambda{\lambda^l\over Q(\lambda)
\sqrt \Delta(\lambda)}=-(t_k-t_k^{(0)})\sum_s{\lambda_s^{k+l}\over \dot Q
(\lambda_s)}\eqno(4.13)$$
 \noindent
In particular, for the  first nontrivial flows ($k=0$ in the DG case and $k=1$
in the DN case) we  get:
$$\sum_r \int_{\mu_r(t^{(0)})}^{\mu_r(t)}d\lambda{\lambda^l\over Q(\lambda)
\sqrt \Delta(\lambda)}=-(t-t^{(0)})\times\pmatrix{ \delta_{l,N-2}~~(DN)\cr
 \delta_{l,N-1}~~(DG)\cr}\eqno(4.14)
$$

The evaluation of the discrete-time evolution of the separation variables
$\mu_j$ is considerably more involved.
The basic starting point is now the discrete Lax equation (2.6a), that entails
both in the DG and in the DN case :

$$L_3^\prime (\mu_j) = -\pi_j \eqno(4.15a)$$

\noindent
to be of course complemented by:

$$L_3^\prime (\mu_j^\prime)= \pi_j^\prime \eqno(4.14b)$$

\noindent
Eq.(4.15a) allows one to express the quantities $\{p_k^\prime q_k^\prime\}$
in terms of $\{\mu_j\},\{\pi_j\}$; by inserting such expression into (4.15b),
one gets a set of formulas relating $\{\mu_j^\prime\},\{\pi_j^\prime \}$ to
 $\{\mu_j\},\{\pi_j\}$ or, in other words, relating $\{\mu_j^\prime\}$
to $\{\mu_j\}$ through the map invariants. The explicit calculation are rather
cumbersome but on the other hand straightforward, relying upon repeated
applications of the residues theorem. Omitting the details, we just report
the final result:

$$\mu_j^\prime-(M-\Lambda)=\epsilon - {2Q(\mu_j^\prime)\pi_j^\prime \over
 P(\mu_j^\prime)}-2\sum_s{1\over \mu_j^\prime-\mu_s}{Q(\mu_s)\pi_s\over \dot
P (\mu_s)}\eqno(4.16)$$

\noindent
where:

\noindent
$\epsilon=1$ and $j=1,\cdots,N-1$ in the Neumann case

\noindent
$\epsilon=0$ and $j=1,\cdots,N$ in the Garnier case

\noindent
$M=\sum_k \mu_k~~;~~\Lambda=\sum_k \lambda_k$
\smallskip

By recalling that $\pi_j={\sqrt{\Delta(\mu_j)}}$ and getting rid of the
square roots, one gets an algebraic equation of degree $2N (2(N-1))$ in the
DG (DN) case, as the coefficients of powers $2N+2, 2N+1 (2N,2N-1)$ identically
vanish. Of course, only half of the roots of that algebraic equation will also
solve (4.16), where a definite sign for the $\pi_j $ has to be chosen.
Once the $\mu_j$ have been found, the frequencies $\nu_j$ of the maps can be
evaluated by hyperelliptic integrals. In fact, denoting by $\theta_j$ the
variable conjugated to $I_j$, given by:

$$\theta_j=\sum_k\int^{\mu_k} {\partial \pi(\mu)\over \partial I_j}d\mu~=
{1\over 2}\sum_k\int^{\mu_k}{1\over(\lambda_j-\mu){\sqrt{\Delta(\mu)}}}d\mu$$

\noindent
we have:
$$\nu_j(I_1,\cdots,I_N)~=~{1\over 2}\sum_k\int_{\mu_k}^{\mu_k^\prime}
{1\over(\lambda_j-\mu){\sqrt{\Delta(\mu)}}}d\mu$$

\noindent
and the maps linearize to:
$$\theta_j(n)~=~\nu_jn+\theta_j(0)$$
\vskip 1 true cm
\noindent
{\bf{Concluding Remarks}}

\medskip
We have shown that the $r$-matrix approach to integrable systems
can be succesfully applied to the discrete time case, starting from
the Lax representation. It remains an open question whether one
 can construct different
Lax pairs, possibly in terms of $N\times N$ matrices with a
 polynomial dependence upon the spectral parameter, leading to constant
$r$-matrices. Another problem worth to be looked at is
the role of the $r$-matrix as far as the discrete-time dynamics is concerned;
in fact, since to an integrable map one can associate a (family of)
interpolating hamiltonian flow(s), engendered by a (family of)
hamilton function(s) functionally dependent on the invariants, one has
to expect that the matrix $A$ entering in the discrete Lax representation
be expressible as well as a function of the matrices $M^{(k)}$ that
define the compatible continuous dynamics and are constructed through the
$r$-matrix. Finally, a few words about quantisation : without going here
into  the stimulating problem  of quantising symplectic
maps, we just mention two points: i) once symmetrized
with respect to $p$ and $q$ the classical invariants (2.7b) (2.11b) have
a simple quantum version to commuting (formally) self-adjoint operators;
ii)  as illustrated in [11]
in the continuous case,  the  separation equations can be quantised to
a set of decoupled one-dimensional multiparametric ordinary
differential equations of Schroedinger type.

\vskip 1 true cm
\noindent
{\bf{Acknowledgements}}

It is a pleasure to ackowledge stimulating discussions with J.Avan
and C.Viallet about dynamical {\it{r}}-matrices. This research has been
partially supported by the Italian M.U.R.S.T., in the context of the
National Research Project ``Problemi matematici della Fisica".
\vfill\eject
\noindent
{\bf{References}}

\item {[1]} Cao Ce Wen, "Classical integrable systems generated through
      nonlinearization of  eigenvalue problems" in {\it{Nonlinear}}
{\it{Physics}},
      Gu C et al. eds., Springer Verlag, Berlin 1990.

\item { } Li Yi Shen and Y Zeng, J.Math.Phys. 30 (1989), 1679.

\item { } M.Antonowicz, S.Rauch-Wojciechowski, Phys.Lett.147A (1990), 455;

\item { } M.Antonowicz, S.Rauch-Woijciechowski, J.Phys.A:Math.Gen 24
(1992),5043.

\item {[2]} G.R.W.Quispel, J.A.G.Roberts, C.J.Thompson,,Phys.Lett.A 126
(1988),419.

\item { } G.R.W.Quispel, J.A.G.Roberts, C.J.Thompson, Physica D 34 (1989),
      183-192.

\item { } H.W.Capel, F.W.Nijhoff, V.G.Papageorgiou,
      Phys.Lett. A147 (1990), 106-114.

\item { } H.W.Capel, F.W.Nijhoff, V.G.Papageorgiou,
      ``Lattice Equations and Integrable Map-

\item { } pings", in {\it{Nonlinear}}
      {\it{Evolution Equations and Dynamical Systems}}, S.Carillo

\item { } and  O.Ragnisco
      eds., Springer Verlag, Berlin 1990, 182.

\item { } M.Bruschi, O.Ragnisco, P.M.Santini
      and Tu Gui-Zhang, Physica D 49 (1991), 273-294.

\item { } J.Moser, A.P.Veselov, Comm.Math.Phys.139 (1991), 217.

\item {[3]} O.Ragnisco, ``A simple
 method to generate integrable symplectic maps", in
 {\it{Soliton and Chaos}}, I.
Antoniou and F.J.Lambert eds.,
      Springer, Berlin-Heidelberg 1991.

\item {[4]} O.Ragnisco, Phys.Lett.A 167 (1992), 165-171.

\item  {[5]} O.Ragnisco, ``Restricted Flows of Toda Hierarchy as Integrable
Maps",
in {\it{Proceedings of the XIX I.C.G.T.M.P.}}, M.Del Olmo and M.Santander
eds., CIEMAT, Madrid 1993.

\item {[6]} Yu. B. Suris, Phys.Lett. A145 (1990) 113

\item {[7]} F.Nijhoff, Gen-Di Pang ``A Time-Discretized
Version of the Calogero-Moser Model",
Preprint Paderborn University 1993.

\item {[8]} Cao Cewen, O.Ragnisco, Yongtang Wu, ``On The Relation of the
Stationary
 Toda
Equations and The Symplectic Maps", submitted to J.Phys.A.

\item {[9]} O.Babelon, C.M.Viallet, Phys.Lett.B 237 (1990), 411-416.

\item {[10]}J.Avan, M.Talon, Intern.Journ.Mod.Phys. 23 (1990), 4477-4488.

\item {[11]} J.C.Eilbeck, V.Z.Enol'skii, V.B.Kuzsnetsov, A.V.Tsiganov, ``Linear
r-matrix algebra for classical separable systems", preprint (1993).

\item { } J.C.Eilbeck, V.Z.Enol'skii, V.B.Kuzsnetsov, D.V.Leykin, Phys.Lett.A
180 (1993) 208-214.

\item {[12]} J.Avan, O.Babelon, M.Talon, ``Construction of the classical
{\it{R}}-matrices for the Toda and Calogero Models", Preprint LPTHE 93-31.

\item { } E.K.Sklyanin, ``Dynamical {\it{r}}-matrices for the Elliptic
Calogero-
Moser Model", preprint LPTHE-93-42.

\item  {[13]} H.Flaschka, Progr.Theor.Phys. 51 (1974), 703.

%\item B.Grammaticos, A.Ramani, V.G.Papageorgiou, Phys.Rev.Lett. 67 (1991),
%1825-27.
  %      \item {[]} F.W. Nijhoff, H.W. Capel and V.G. Papageorgiou,
   %     Phys.Rev. A46 (1992) 2155;
%\item { } F.W. Nijhoff,  V.G. Papageorgiou, H.W. Capel, G.R.W.Quispel,
 %Inv.Prob. 8 (1992) 597;

\vfil\end